\documentclass[aps, prb, superscriptaddress, twocolumn]{revtex4-1}
\usepackage{bm}
\usepackage[fleqn]{amsmath}
\usepackage{hyperref}
\usepackage{graphicx}

\begin{document}

\title{Resonance Absorption of Terahertz Radiation
in Nanoperforated Graphene}

\author{V.V. Enaldiev}
\email{vova.enaldiev@gmail.com}
\affiliation{Kotelnikov Institute of Radio-engineering and Electronics of the Russian Academy of Sciences, 11-7 Mokhovaya St, Moscow, 125009 Russia}

\author{V.A. Volkov}
\affiliation{Kotelnikov Institute of Radio-engineering and Electronics of the Russian Academy of Sciences, 11-7 Mokhovaya St, Moscow, 125009 Russia}
\affiliation{Moscow Institute of Physics and Technology, Dolgoprudny 141700, Russia}

\date{\today}

\begin{abstract}

Recent measurements of the conductivity of nanoperforated graphene are interpreted in terms of edges states existing near the edge of each nanohole. The perimetric quantization of edge states should result in the formation of a quasi-equidistant ladder of quasistationary energy levels. Dirac fermions filling this ladder rotate about each nanohole in the direction determined by the valley index. It is shown that the irradiation of this
system by circularly polarized terahertz radiation leads to a resonance in absorption in one of the valleys. The magnitude of absorption at the resonance frequency can be controlled by means of gate voltage.

\end{abstract}

\maketitle

\section{Introduction}

Graphene has been actively studied both theoretically and experimentally for more than ten years. These studies are stimulated primarily by a specific ultra-relativistic law of dispersion of charge carriers (so-called Dirac fermions), which is responsible for unique properties of graphene. Owing to a large absorption coefficient of graphene in the optical range ($\sim 2.3\%$ ) [\onlinecite{Nair_2008},\onlinecite{Li_2008}], graphite flakes with a thickness of several atomic layers (multilayer graphene) can be identified with an ordinary optical microscope [\onlinecite{Novoselov_Science_2004}]. Absorption in graphene has two contributions [\onlinecite{Gusynin_2006}]: interband and intraband ones. The former contribution is due to vertical interband transitions allowed by the Pauli exclusion principle. It has a universal value [\onlinecite{Gusynin_2006,Falkovsky_2007,Falkovsky_2012,Ando_2002}] independent
of the parameters of graphene. The interband contribution controlled by the gate voltage determines absorption in the infrared and visible ranges. The intraband contribution is crucial in the response of graphene to far infrared radiation and is described within the Drude model [\onlinecite{Horng_2011}]. In this work, we analyze a new mechanism of intraband absorption associated with resonance transitions between edge states. 

An important feature of graphene is the existence of intrinsic edge states [\onlinecite{Ritter_2009},\onlinecite{Tao_2011}]. The existence of dispersionless edge states at a zigzag edge, as well as the absence of edge states at an armchair edge, is predicted by the tight-binding theory in the nearest neighbor approximation [\onlinecite{Brey_2006}]. However, the inclusion of contributions next to nearest neighbors [\onlinecite{Peres_2006}], reconstruction [\onlinecite{Ostaay_Akhmerov_PRB}], or chemical absorption [\onlinecite{Fujii_2014}] at the linear zigzag edge results in the appearance of nonzero dispersion of edge states. The appearance of edge states at a linear reconstructed armchair edge was also predicted [\onlinecite{Maksimov_2013}].

\begin{figure}
\begin{center}
\includegraphics[width=7cm]{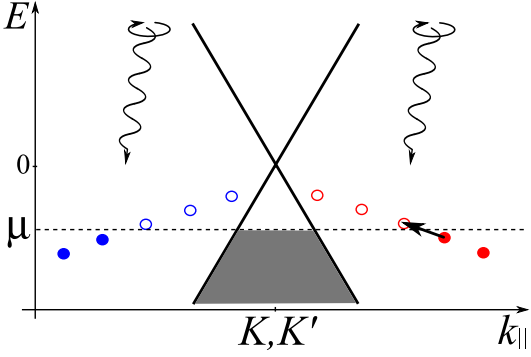}
\caption{\label{Fig2} Semiclassical dependence of the energies of quasistationary edge states in graphene with one antidot on the quantized tangential component of the
quasimomentum ($k_{||}=l/R$ ) in the reduced valley scheme at $a<0$. The red and blue circles are quasistationary levels in the valleys $K$ and $K'$, respectively. The gray background and closed circles are the filled delocalized and edge states located under the Fermi level $\mu$, respectively. Clockwise circularly polarized radiation is responsible for transitions with the change $l\to l-1$ ; for this reason, only the transition (shown by a thick arrow) between edge states from the
red valley results in the resonance in absorption. } 
\end{center}
\end{figure}

However, from the experimental point of view, the situation is much less clear and should be discussed separately. Nevertheless, recent experiments showed [\onlinecite{Latyshev_SciRep_2014},\onlinecite{Latyshev_2013}] that edge states exist at a round edge in nanoperforated graphene. Nanoholes in graphene, which are also called antidots, are regions in a graphene sheet that are unavailable for Dirac fermions. An antidot can be described in the envelope function approximation by a phenomenological boundary condition under which the Hamiltonian of confined graphene is Hermitian:
\begin{equation}\label{eq:graph_Ham}
v\bm{\sigma}\cdot\hat{{\bf p}}\psi_{\tau} = \varepsilon\psi_{\tau},
\end{equation}
where $v\approx 10^6$ m/s, $\bm{\sigma}=\left(\sigma_x,\sigma_y\right)$ is vector of the Puali matrices, $\hat{{\bf p}} =(\hat{p}_x, \hat{p}_y)$ is the momentum operator, $\tau=\pm 1$ is the index of $K$, $K'$ valleys, respectively, $\psi_{\tau} = (\psi_1, \psi_2)^{T}$ are the two-component envelope functions in the valley with the index $\tau$. The general
boundary condition [\onlinecite{Ostaay_Akhmerov_PRB,Volkov_2009,Tkachov_2009,Basko_2009}] that conserves time reversal symmetry and does not entangle valleys at an edge is parameterized by only one real phenomenological parameter $a$:
\begin{equation}\label{main_BC}
\left [\psi_1 + i\tau a^{\tau}e^{-i\varphi}\psi_2 \ \right ]_{edge} = 0,
\end{equation} 
here $\varphi$ is an angle between $x$ axis and unit normal to the edge. Owing to the boundary condition given by Eq. (\ref{main_BC}) and to the finiteness of the perimeter of the antidot, almost equidistant levels of quasistationary edge states corresponding to the rotation of Dirac fermions about the antidot appear. The low-energy spectrum
of edge states at the antidot is given by the expression (see Fig. \ref{Fig2}) [\onlinecite{Zag_Dev_En_PRB}]:
\begin{equation}\label{edge_spectrum}
\varepsilon_{l\tau} = \tau l \left(\hbar\omega_0\,{\rm sgn} a - \Delta_l\right) - i\gamma_{l}.
\end{equation}
Here, $l$ is the orbital angular momentum of an edge state; the equidistant part of the spectrum is determined by the frequency $\omega_0 = 2|a|v/R$, where $R$ is the radius of the antidot, and a small non-equidistant correction has the form $\Delta_l = [\hbar v/R] 2 a^3l(1-l\delta_{|l|1})/(l-1)$, where $\delta_{|l|1}$ is the Kronecker delta and $a$ is the parameter of the boundary condition (\ref{main_BC}) averaged over the perimeter. A small finite inverse lifetime of the edge state is:
\begin{equation}\label{peak_wide}
\gamma_{l} = \frac{2\pi\hbar v}{R}\frac{(|al|)^{2|l|+1}}{\Gamma(|l|)\Gamma(|l|+1)} ,
\end{equation}
where $\Gamma(l)$ is the Gamma function. The region of applicability of the spectrum of edge states (\ref{edge_spectrum}) is determined by the conditions $\varepsilon_{l\tau} R/\hbar v\ll 1, |al|\ll 1, \tau l>0, |l|=1,2,3,\dots$, under which the energies of
edge states are well defined in view of ${\rm Re} \varepsilon_{l\tau}\gg {\rm Im} \varepsilon_{l\tau}$.

In this work, we show that the absorption coefficient of nanoperforated graphene has a resonance at frequencies corresponding to the difference between the energies of the nearest levels (\ref{edge_spectrum}). Absorption at the resonance frequency controlled by the gate voltage can reach several percent at a low but experimentally achievable concentration of antidots $n_a\ll 1/R^2$.

\section{RESONANCE ABSORPTION IN NANOPERFORATED GRAPHENE}

We construct the complete system of functions for infinite graphene with one antidot of the radius $R$. Since we are interested only in intraband absorption, a band with negative energies $\varepsilon_k$ is considered for definiteness. It is convenient to introduce the system of functions of the scattering problem that has a certain asymptotic behavior at infinity, where there are a plane wave $e^{i{\bf kr}}$ with the wave vector ${\bf k}=[|\varepsilon_k|/\hbar v](\cos\vartheta,\sin\vartheta)$ and a cylindrical wave divergent from the antidot. The exact wavefunction of the scattering problem in cylindrical coordinates has the form
\begin{eqnarray}\label{continius_spectra_wf+}
\psi_{{\bf k}}^{(+)}=\qquad\qquad\qquad\qquad\qquad\qquad\qquad\qquad\qquad\qquad \nonumber \\ =\frac{1}{\sqrt{2}}\sum_{l=-\infty}^{+\infty}
\left (
\begin{array}{l}
J_l(kr)+C_{l\tau}(k) H_{l}^{(2)}(kr)  \\
-i\left[J_{l+1}(kr)+C_{l\tau}(k) H_{l+1}^{(2)}(kr) \right]e^{i\varphi}
\end{array} 
\right )\times\nonumber \\
\times i^{l}e^{-i\vartheta l+il\varphi}.
\end{eqnarray}
Here, $k=|\varepsilon_k/\hbar v|$, the terms with the Bessel function $J_l$ describe the expansion of the plane wave in functions with the orbital angular momentum $l$, and the terms with the Hankel function $H^{(2)}_l$ represent the divergent cylindrical wave. The coefficients $C_{l\tau}(k)$ are determined from the boundary condition (\ref{main_BC}) as
\begin{equation}\label{coefficients_in_scat_amplitude_valley_1}
C_{l\tau}(k) = -\frac{J_l(kR)+\tau a^{\tau}J_{l+1}(kR)}{H_{l}^{(2)}(kR)+\tau a^{\tau}H_{l+1}^{(2)}(kR)}.
\end{equation} 
In addition to functions (\ref{continius_spectra_wf+}), the calculation of matrix elements of transitions requires the functions $\psi_{{\bf k}}^{(-)}$ corresponding
to the plane wave $e^{i{\bf kr}}$ at infinity and a cylindrical wave converging to the antidot [\onlinecite{LL_2008_english}]. They are obtained from Eq. (\ref{continius_spectra_wf+}) by the substitution $C_l\to C_l^{*}$ and $H^{(2)}\to H^{(1)}$, where $H^{(1)}$ and $H^{(2)}$ are the Hankel functions of the first and second kinds, respectively. Both systems of functions $\psi_{{\bf k}}^{(+)}$ and $\psi_{{\bf k}}^{(-)}$ are complete and normalized to the delta function:
\begin{equation}\label{norma}
\int_{R}^{+\infty}rdr\int_{0}^{2\pi}d\varphi \psi_{{\bf k}}^{(\pm)+}\psi_{{\bf k'}}^{(\pm)} = (2\pi)^{2}\frac{\delta(k-k')}{k}\delta(\vartheta  -\vartheta'). 
\end{equation}

Let weak clockwise circularly polarized radiation whose electric field has the form ${\bf F} = F(\cos\omega t, -\sin\omega t, 0)$ be normally incident on graphene. Introducing the interaction of Dirac fermions with radiation in terms of the vector potential ${\bf A} = -c\int{\bf F}dt$, we obtain the following term describing transitions in the continuous spectrum:
\begin{equation}\label{perturnbation_in_Hamiltonian}
\begin{array}{l}
V = v\frac{e}{c}{\bf \sigma}\cdot{\bf A}= \frac{eFv}{2\omega}(-i\sigma_x-\sigma_y)e^{i\omega t} + \frac{eFv}{2\omega}(i\sigma_x-\sigma_y)e^{-i\omega t} \\
\equiv V_1e^{i\omega t} + V_2e^{-i\omega t}
\end{array}
\end{equation}
Here, $V_1$ and $V_2$ describe the emission and absorption of a photon. The probability of transition per unit time is given by the formula
\begin{equation}\label{transition_probability}
dw^{(1,2)}_{{\bf k}{\bf k'}} = \frac{2\pi}{\hbar}\left|\langle {\bf k'}|V_{1,2}|{\bf k}\rangle\right|^2\delta(\varepsilon_{k'}-\varepsilon_{k} \pm \hbar\omega)\frac{d^2k'}{(2\pi)^2},
\end{equation} 
where the sign $+(-)$ responds to the superscript 1(2),  $|{\bf k}\rangle\equiv \psi_{{\bf k}}^{(+)} $, $\langle{\bf k'}|\equiv \psi_{{\bf k'}}^{(-)\dag} $, the matrix element is determined by the expression:
\begin{equation}\label{matrix_element}
\begin{array}{l}
\langle {\bf k'}|V_1|{\bf k}\rangle = 
\dfrac{4evF e^{i\vartheta}}{\omega(k^2-k'^2)\pi R^2kk'}\times\\ \sum_{l}\frac{e^{i(\vartheta'-\vartheta)l}\left (kR+\tau 2a^{\tau}(l+1)+a^{2\tau}k'R\right )}{\left[H_{l}^{(2)}(k'R)+\tau a^{\tau}H_{l+1}^{(2)}(k'R)\right ]\left[H_{l+1}^{(2)}(kR)+\tau a^{\tau}H_{l+2}^{(2)}(kR)\right]},
\end{array}
\end{equation}  
and the matrix element $\langle {\bf k'}|V_2|{\bf k}\rangle$ is obtained from
Eq. (\ref{matrix_element})  by substituting $k\leftrightarrow k'$.

\begin{figure}[ht]
 \includegraphics[width=8cm,height=6cm]{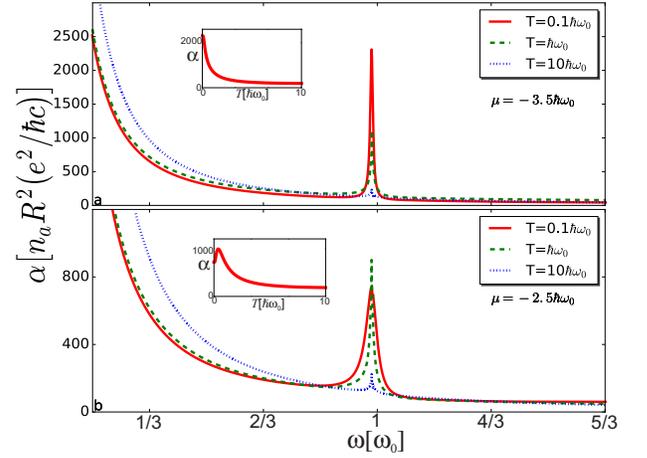}
 \caption{\label{Fig1}  Frequency dependence of the absorption coefficient given by Eq. (\ref{absorbtion_exact}) at $a=-0.15$ for the Fermi level (a) $\mu = -3.5\hbar\omega_0$ and (b) $\mu = -2.5\hbar\omega_0$ . According to Eq. (\ref{res_freq}), the resonance frequency is $\omega_{res} \approx 0.99 \omega_0 \approx 5$ THz at $R=10$ nm. The insets show the temperature dependence of the resonance amplitude for the respective Fermi levels.}
\end{figure}

We define the dimensionless absorption coefficient $\alpha(\omega)$ as the power dissipated by a unit area of graphene with antidots normalized to the total incident flux
\begin{equation}\label{absorbtion}
\alpha(\omega) = \frac{U(\omega)}{\Omega S},
\end{equation}
where $\Omega$ is the area of the graphene sheet, $S=cF^2\sqrt{\kappa}/4\pi$ is the magnitude of the Poynting vector of incident radiation, and $\kappa$ is the effective relative permittivity of the system. The dissipated power is expressed in terms of the transition probabilities as
\begin{eqnarray}\label{dissipation_power}
U(\omega) = N_a\hbar \omega \int\int\left [ \frac{d^2k}{(2\pi)^2}dw^{(2)}_{{\bf k}{\bf k'}}f({\bf k})(1-f({\bf k'}))- \right. \nonumber\\ \left. - \frac{d^2k'}{(2\pi)^2}dw^{(1)}_{{\bf k'}{\bf k}}f({\bf k'})(1-f({\bf k})))\right ],
\end{eqnarray}
where it is assumed that contributions from all antidots are independent, $N_a$ is the number of antidots, and is the Fermi--Dirac function. Formula (\ref{dissipation_power})
is valid at a low concentration of antidots $n_a R^2 \ll 1$ ($n_a= N_a/\Omega$). The substitution of Eq. (\ref{dissipation_power}) into Eq. (\ref{absorbtion}) gives the following final form for the intraband contribution to the absorption coefficient of perforated graphene:
\begin{equation}\label{absorbtion_exact}
\begin{array}{l}
\alpha(\omega) = \frac{32g_sn_ae^2v^2\left(\frac{\hbar v}{R}\right)^2}{\pi^2\omega c } 
\int_0^{+\infty}d\varepsilon\left\{\frac{f(-\varepsilon-\hbar\omega)-f(-\varepsilon)}{\varepsilon(\varepsilon + \hbar\omega) ((\varepsilon + \hbar\omega)^2-\varepsilon^2)^2}\right\}\times \\
\\
\sum_l\frac{\left ( \varepsilon+\hbar\omega + \tau 2a^{\tau}\hbar v(l+1)/R + a^{2\tau}\varepsilon\right )^2}{\left|H_{l}^{(2)}(kR)+\tau a^{\tau}H_{l+1}^{(2)}(kR)\right|^2\left|H_{l+1}^{(2)}(kR+\omega R/v)+\tau a^{\tau}H_{l+2}^{(2)}(kR+\omega R/v)\right|^2},
\end{array}
\end{equation}
where $g_s = 2$ is the spin degeneracy factor in graphene. The frequency dependence of absorption is shown in Fig. {\ref{Fig1}} for various temperatures and Fermi levels. Both denominators in the sum over $l$ in Eq. (\ref{absorbtion_exact}) are small when the energy $\varepsilon_k$ of the state $|{\bf k}\rangle$ corresponds to the real part of the ($l+1$)-th quasistationary edge level (\ref{edge_spectrum}) and, simultaneously, $\varepsilon_k+\hbar\omega$ is equal to the real part of the energy of the $l$-th quasistationary level. In fact, this is the condition for a resonance in
absorption. For the considered polarization (clockwise)
and valence band, the indicated condition is satisfied
only in one of the valleys $\tau=+1$ at $l>1$ and $a<0$ because the energies of quasistationary edge states in this valley decrease with an increase in $l$ (see
Fig. \ref{Fig2}). At the opposite polarization, the resonance in
absorption appears when $\varepsilon_k$ corresponds to the energy
of the ($l-1$)-th level of edge states and $\varepsilon_k+\hbar\omega$ coincides
with the energy of the $l$-th level of edge states. This condition is satisfied only in the valley $\tau=-1$ at $l<0$ and $a<0$ because change in the direction of
polarization is equivalent to time reversal connecting two valleys. In the limit $\omega\to 0$, the absorption coefficient (\ref{absorbtion_exact}) follows the Drude behavior $\omega^{-2}$ in the pure system ($\omega\tau=\infty$ ).

\begin{figure}
 \includegraphics[width=8cm]{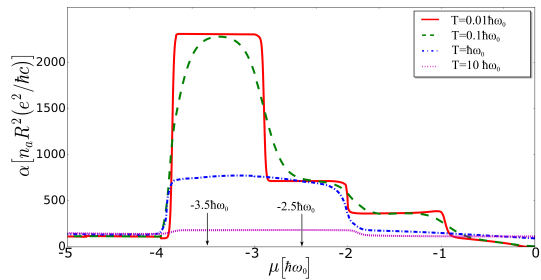}
 \caption{\label{Fig3} Absorption at the resonance frequency $\omega_{res} = 0.99 \omega_0$ versus the position of the Fermi level at $a=-0.15$. }
\end{figure}

In the low-energy limit $kR\ll 1$, it is convenient to represent the absorption in the form
\begin{equation}\label{absorption_low_energy}
\alpha(\omega)=\sum_{l=1}^{l_{max}}\alpha^{(res)}_l(\omega, T) + Z(\omega,T),
\end{equation}
where $Z(\omega,T)$ is the smooth function of the frequency and the partial resonance term is given by the expression 
\begin{eqnarray}\label{l_resonanse}
\alpha^{(res)}_{l}(\omega,T) =\frac{\sinh\left(\frac{\hbar\omega}{2T}\right )}{\cosh(\frac{Re(\varepsilon_{l+1})-\mu}{T}+\frac{\hbar\omega}{2T}) + \cosh\left(\frac{\hbar\omega}{2T}\right )}\times\qquad\nonumber \\ 
\times\frac{8\sqrt{\pi}g_sn_ae^2v^2}{\omega c }\frac{(l+1)a^2\gamma_{l}}{l\left [\hbar^2(\omega- \omega_l)^2+\gamma_{l}^2\right ]}.\qquad
\end{eqnarray}
Here, the resonance frequency has the form
\begin{equation}\label{res_freq}
\omega_{l} = {\rm Re}\left( \varepsilon_{l} - \varepsilon_{l+1}\right)/\hbar =\omega_0 \left[1 - a^2\left(1+4\delta_{l1}\right)\right]
\end{equation}
Formulas (\ref{l_resonanse}) and (\ref{res_freq}) are applicable at $|al|\lesssim 1$; this condition provides the estimate of the number of resonance terms in Eq. (\ref{absorption_low_energy}): $|l_{max}|\sim 1/|a|$ . Therefore, the observation of resonances is possible only when the position of the Fermi level with respect to the Dirac point satisfies the condition $|\mu|\lesssim \hbar\omega_0/2|a| $. According
to Eq. (\ref{l_resonanse}), the contribution to the sum from the $l$-th term at the resonance frequency is proportional to the lifetime of the quasistationary state ($\propto 1/\gamma_l $), which increases with $l$ according to Eq. (\ref{peak_wide}). Figure 3 shows the dependence of the absorption coefficient at the resonance frequency on the position of the Fermi level for four characteristic temperatures. It is seen that
absorption at low temperatures is a step function of the position of the Fermi level. The resonance amplitude at low temperatures ($T\ll \hbar\omega_0 $) is determined by a
term $\alpha^{(res)}_{l}$ in sum (\ref{absorption_low_energy}) for which $f(Re(\varepsilon_{l}))\approx 0$ and $f(Re(\varepsilon_{l})-\hbar\omega)\approx 1$. The contribution of terms with large values to the absorption at the resonance frequency
increases with the temperature. In particular, the resonance in Fig. \ref{Fig1}b at the temperature $T=0.1\hbar\omega_0$ is determined by the term with the number $l=2$
in Eq. (\ref{absorption_low_energy}). At an order of magnitude higher
temperature ($T=\hbar\omega_0$), an additional contribution comes from the term with $l=3$ , which results in an increase in the resonance maximum and in a decrease in its width (because $\gamma_{l} \gg\gamma_{l+1}$). A further increase in the temperature leads to the reduction of the maximum value (see the inset of Fig. \ref{Fig1}b). A different situation is seen in Fig. \ref{Fig1}a, where the leading contributing
to absorption at the resonance frequency at temperatures much lower than the resonance energy comes from the term with the number $l=3$ in sum (\ref{absorption_low_energy}) (at $\mu=-3.5\hbar\omega_0$). Its contribution to absorption is leading at the considered value $a=-0.15$ because it is triple the contributions from the other terms (see Fig. \ref{Fig3}). For this reason, the width of the resonance line is independent of the temperature and the maximum of this line decreases with increasing temperature (see the inset of Fig. \ref{Fig1}а). Thus, the difference between temperature
dependences of the resonance amplitude in Figs. \ref{Fig1}a and \ref{Fig1}b is due to the competition between two factors. On the one hand, with an increase in the temperature, the difference between the Fermi distribution functions decreases and absorption decreases. On the other hand, additional contributions to the resonance
amplitude appear because of neighboring resonance transitions. The appearance of monotonic and nonmonotonic temperature dependences of the resonance amplitude is determined in a complicated way by the parameters of the system. 
The dependence of the absorption at the resonance frequency on the position of the Fermi level shown in Fig. \ref{Fig3} demonstrates the possibility of controlling the response at the resonance by means of gate voltage. At low temperatures and experimentally achievable concentration of antidots  $n_aR^2=2\cdot 10^{-3}$ [16], the absolute value of the absorption coefficient can reach several percent (for $a=-0.15$ and $-4\hbar\omega_0 \lesssim \mu\lesssim -3\hbar\omega_0$), which is comparable with the plasmon response of graphene structures [\onlinecite{Ju_2011}]. The absorption coefficient at high temperatures ($T\gg \hbar\omega_0,\mu$) decreases as $1/T$ (see the insets of Fig. \ref{Fig1}).

\section{Conclusions}

It has been shown that the intraband part of the response of Dirac fermions in nanoperforated graphene to external electromagnetic radiation has a resonance caused by transitions between the nearest levels of edge states existing near antidots. For antidots with a nanometer diameter, the resonance lies in the terahertz spectral range. Circularly polarized radiation gives the resonance only in one of the valleys. Absorption at low temperatures is a step function of the Fermi level position. Absorption at the maximum can be controlled by varying the concentration via the gate
voltage, which makes it possible to use nanoperforated graphene as an optical modulator for the terahertz range [\onlinecite{Sun_2016}]. 

This work was supported by the Russian Science Foundation (project no. 16-12-10411).

\bibliography{thesis}

\end{document}